# Geographical Asynchronous Information Access (GAIA) in the Cloud

*Ali Elouafiq, Mohamed Riduan Abid*
*Alakhawayn University in Ifrane, Morocco*
{A.Elouafiq,R.Abid }@aui.ma

**Abstract**

Non-relational databases are the common means of data storage in the Cloud, and optimizing the data access is of paramount importance into determining the overall Cloud system performance.

In this paper, we present GAIA, a novel model for retrieving and managing correlated geo-localized data in the cloud environment. We survey and compare the existing models used mostly in Geographical Information Systems (GIS), mainly the Grid model and the Coordinate's Projection model. Besides, we present a benchmark comparing the efficiency of the models.

Using extensive experimentation, we show that GAIA outperforms the existing models by its high efficiency which is of $O(\log(n))$, and this mainly thanks to its combination of projection with cell decomposition. The other models have a linear efficiency of $O(n^2)$. The presented model is designed from the ground up to support GIS and is designed to suit both cloud and parallel computing.

## I. Introduction

Current Geographical Information System (GIS) and Geographically Correlated Data (GCD) play an important role in the social network realm [1], especially with the latest widespread usage of geo-localized information for mobile services. During the deployment of a real-world mobile social network (jabeklah) [1], we have encountered some performance and QoS issues that are directly related to geo-localized data management in the back-end servers. In a cloud environment [4], the cost of hosting and managing an application are billed based on the CPU and memory usage consumed by the application. In a social networking business the response delay of the application is a key metric for measuring user satisfaction. Therefore optimizing both the average time delay and computation time at the server level becomes a crucial need. To further highlight the issue, in any web service that is based on geo-localized customization, there are many users requesting similar services in the same region at the same time. As an example, assuming we have 10 users in a 500 meter range that requires a specific service at a time. The application should not talk to the non-relational database (NRDBs) in multiple rounds, but rather should be efficient enough by grouping the similar queries, based on their location, and optimize its processing to give a quicker response with a lower cost. Henceforth, we need to find a model that permits the access of geographically correlated data directly (random access) and asynchronously.

To tackle this issue, we started first by reviewing the relevant literature, i.e., in the topics related to GIS issues, spatial data management, and cloud computing. We found significant works relevant to the field of implementing GIS in cloud computing. Mainly most of the work highlights software architecture [2], models for applications [3], software-as-a-service applications [4], and deployment of existing GIS technologies in infrastructure-as-a-service environment [5], but they did not tackle the issues related to the use of parallel-architectures and high-replication databases (HRD) with GIS. To address this matter, we have gone through the models used to build GIS and spatial data management systems. Mainly two models are used by most algorithms, systems, and architectures: the Grid model and the projection model.

After comparing the existing information, we have performed a deep mean-to-end decomposition analysis in order to generate a new model that fits the challenging requirements of cloud computing. In this scope, we present the Geographical Asynchronous Information Access (GAIA), an original model based on two main design components: a mathematical transformation (hashing function) and an asynchronous data access algorithm. The asynchronous data access algorithm was designed based on the hashing function to enable direct access of correlated data. The hashing function itself is a merge of cellular decomposition (grids) [6] and coordinate projection to gives one dimensional partial key to access the desired data.

In order to evaluate our model, a benchmark was conceived specifically for this, which revolves around the evaluation of the average time delay (ADT) per query. The benchmark goes through three evaluation criteria, which are single query evaluation, concurrent queries evaluation, and performance uniformity evaluation. At the end of the benchmark, a comparative table is introduced to synthetize the test results. Experimental results conducted in this study showed that GAIA by far outperformed the existing models.

The rest of the paper is organized as follows: section 2 details the related work including GIS in cloud computing and the spatial correlation fundamental models. Section 3 proposes the GAIA model. Section 4 examines benchmark evaluation criteria to evaluate the ADT of each model. Section 5 presents experimental results and observations. Section 6 gives the conclusions and directions for future research.

## II. Related Work

Recently, a number of enterprises and organizations have started investigating and developing technologies and infrastructure for cloud computing [7]. Besides, for decision making purposes, businesses and international organizations need information about trends in different regions in the world, tracks of progresses geographically, and other types of information that is correlated with its location.

In the industrial cloud computing, Amazon Elastic Compute Cloud (EC2) [5] provides a virtual computing environment that enables a user to create his own machines. With the same provider, Amazon SimpleDB [5] is a web service providing the core database functions of data indexing and querying. This service works in close conjunction with Amazon Simple Storage Service (Amazon S3) [5] which provide the ability to store, process, and query data. Google App Engine (GAE) [2] and Microsoft Azure [2] are platforms for developing and hosting web applications in a Platform as a Service schema in (PaaS) [2] . For GAE, Google High Replication Data Store (HRD) [8] is the commonly used schema for data storage. As many other High Replication Databases, Google's HRD is implemented in a Non-Relational Database schema, implemented using Google Big Tables Architecture [8]. In the Open Geospatial Consortium (OGC) [7], a non-profit international consortium of 458 companies, industries, and governments, are developing publically available standards. Still, many issues have been raised about OGC6/7 (Web Service Phase), which have an emphasis on cloud computing.

In the scope of cloud computing, in the Chinese Academy of Science, the Remote Sensing team [2] designed management architecture for spatial information. The study focuses on using the cloud computing for GIS. The spatial information system based on the cloud they have presented revealed an interesting way of adding a GIS layer to the Software-as-a-Service (SaaS) application, without interfering with low-level interventions in the Data Storage. In the Cloud Computing and Distributed Systems Laboratory in University of Melbourne [5], a cloud computing oriented GIS architecture has been designed, which enables the deployment of standard GIS systems in existing Amazon Web Services (AWS) technologies. However, in the existing work, efforts have focused on the deployment of GIS in cloud environments or the design of architectures that suits geo-spatial applications, but none had targeted the design of a data access and management model adapted for the nature of non-relational databases in a cloud environment.

In order to access GCD storage must take into account the spatial coordinates of the store data. We will cover in this section the existing methods of accessing GCD. Either using spherical, Euclidian, or polar coordinates, GCD are referenced using 2D objects [6]. In other words, GCD is referenced using two variables. With this approach, the use of R-Trees and Quad trees at the low-level of the database can support the categorization, and smart retrieval of the data [9]. However, this method of storage raises many issues in the NRDB environments: Due to its 2D nature and to the large size of HRDs, correlated data are difficult to fetch, which requires additional computational costs at the level of the application side. In small regions of 2D data, projection from 2D to 1D is used to store GCD in 1D table [10]. This can be seen as the coordinates are hashed into a single variable which enables them to be correlated according to their hashes. In other words, in a specific region of the 1D table for each datum there is at least one datum that is correlated to it. The problem with this approach persists in the collisions. Each 2D-to-1D mapping always lead to a large amount of collisions based on the projection reference. This can be seen as chunking the region into smaller HRDs, raising again the issues of the Plain Coordinate [10] for filtering the GCD. This will result in having non-useful accessed data for the GCD.

Inspired from metrological applications and the 3D computational graphics realm [4], the grid referencing [4] can be simply explained as partitioning the longitude-latitude surface into a grid. Each cell can be seen as a region, which makes correlated data gathered in the same cell or neighbour cells. This approach is the most used in GIS systems nowadays; it has proven its efficiency and adaptability to the relational databases and R-Tree and Quad-Tree implementations [3], as well as managing shapes and polygon data structures in the GIS. Nevertheless, with the growing issue of HRDs for their non-relational nature and the huge data they hold, GCD access and management issues are the same as Plain Coordinates Referencing [9], and is just scaled down with a proportion, even if it remains the most used method for GCD.

To sum up, various works have been conducted to implement GIS systems in cloud environments. However, most of the work covered how to integrate existing GIS systems with existing cloud environments, or how to add GIS layers to SaaS applications .Also, the existing data access and management models are not suited for the nature of data storage and access in the cloud environments. Henceforth, GAIA was designed both to fit the cloud computing nature and to adapt to parallel computing.

## III. GAIA: Geographical Asynchronous Information Access

In order to get full benefits of the existing methods, we have gone through the chained mean-to-end analysis [REF]. This is to extract the advantages of each method and avoid their constraints.

Accounting for to the nature of Non-Relational Databases data, the following are the main requirements that drove our model design:
- Correlated data should be near each other in NRDBs.
- Correlation is represented in the data storage.
- Data needs to be accessed simultaneously.
- Correlated data can be fetched in a limited call series.
- Using asynchronous non-concurrent queries, correlated data can be accessed in one call and in maximum constant time t.

## III.1 Mathematical Transformation

In this model, we represent the longitude and latitude as a 2D rectangle, and any location/point is represented with a *p(x,y)* coordinates within the rectangle. Since the space is limited, with the exact partitioning, the rectangle can be projected into a one dimensional line (Fig.2). The rectangle is partitioned into small squares (Fig.1), and each square represents a cell. This latter is represented as "*c*". Thus, the rectangle has the following characteristics:
- minD: the leftmost x coordinate,
- maxD: the rightmost x coordinate.
- minH: the lowest y coordinate
- maxH: the highest y coordinate.
- D= minD –maxD: width of the rectangle
- H= minD-maxD: height of the rectangle
- c: he cell side.
- $d = D/c$ : the discrete width
- $h = H/C$ : the discrete height

In Fig. 1, a geo-disc Grid [4] is showed, and the queried region is represented by the red circle. The cells where the information resides are colored in yellow. The rectangle has the property of being continuous; using the cell decomposition we can project the rectangle into a discrete line, See Fig. 2, and using the following transformation, that is transforming the location p(x,y) into their respective cell numbers, through transforming continuous coordinates to their discrete cell coordinates:

$$C(p) = (C(x), C(y)) = \left(\frac{x}{c}, \frac{y}{c}\right) \quad (1)$$

Transforming the rectangular cells into their respective linear cells (Fig. 2), so the hash transformation of a point p(x,y) will be:

$$H(p) = \frac{x + y*D}{c} \quad (2)$$

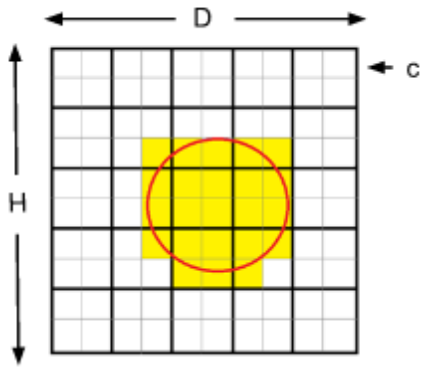

*Fig. 1 Representation of correlated data inside the disc*

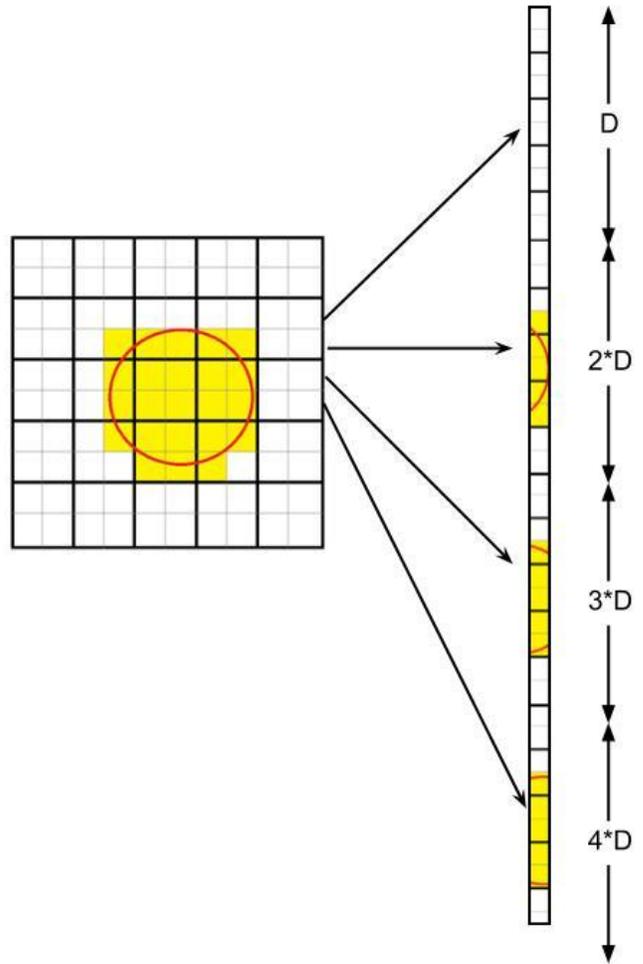

*Fig.2 Projection of the 2D space into 1D discrete list*

In the next section, we will cover an indexing algorithm that will be used to query the different regions of the non-relational database. Thus, this model acquires the precision of the Fisher Linear Discriminant and the correlation of the Geo-Disc cells, minimizing the query time and the unnecessary data.

## III.2 The Geographical Asynchronous Information Access (GAIA) Algorithm

GAIA uses the hash transformation to retrieve the data stored in the cloud NRDBs. This is done in a parallel and an asynchronous mode. Using the hash function, data that is related to each other is indexed easily within the database using the hash transformation, and can be segmented into a limited number of segments in the table that can be retrieved asynchronously.

At the start, the algorithm gets the shape of the geo-area to process; it can be any geometrical shape definition (circles, squares, or polygons) that defines the geographical correlation of data [9]. Then, in order to process the geo-area, using the hash transformation of mathematical model we described earlier, the algorithm divides the shape into a limited number of segments *"S"*. The hash transformation generates a list of couples *(min[i],max[i])*, where each *min[i]* and *max[i]* are respectively the start and the end of the segment number *"i"*. The couple *(min[i],max[i])* is used to query specific regions in the NRDB. Consequently, the segments can be fetched asynchronously or in parallel from the NRDB. Once the queries are done, the data retrieved is geographically correlated to the specified geo-area (See Algorithm in Fig.3).

> *BEGIN*
> - Get the geometrical shape of the geo-area to process
> - Divide the shape into *Limited Segments* (max[i],min[i])
> - Fetch in parallel the data from each (min[i],max[i]) in the NRDB
> - Gather fetched data
> - Return results
>
> *END*

*Fig.3 the GAIA Algorithm*

To have a clearer view on the algorithm, we are presenting an example using the disc as a geo-area (See Algorithm in Fig.4). We suppose that the area inside the disc is the area that represents the correlation of the geo-located data. For a disc shape definition, we take into consideration two main variables, the radius *R*, and the disc centre *p(x,y)*, and a constant parameter *"c"* which represents the cell side used by the implementation of the mathematical model. According to the mathematical transformation the *"i"* will vary between $(p_Y - R)/c$ and $(p_Y + R)/c$ (the width of the disc). Consequently the limiters of each segment *min[i]* and *max[i]* will be respectively $H(p_X - R, i*c)$ and $H(p_X - R, i*c)$. The function *"AsynchronousGetRow"* is an asynchronous request that runs in the background and the execution of next requests does not depend on it. Thus, running asynchronous data fetch will enable us to fetch correlated data in parallel from the NRDB, afterwards the data gathered is returned after waiting for each request to finish.

> *BEGIN*
> - Get R, $p_X, p_Y$, c
>   // This loop goes over all the segments
> - For i ← $(p_Y-R)/c$ to i ← $(p_Y+R)/c$:
>   //Divides the shape into Limited Segments
>   //(max[i],min[i])
>   ○ Data[i]← AsynchronousGetRow[$H(p_X-R, i*c), H(p_X-R,i*c)$]
> - Wait For All Requests to Finish
> - Gather fetched Data[]
> - Return results Data[]
>
> *END*

*Fig.4 GAIA Algorithm using disc geo-area correlation*

## IV. Benchmark

In order to benchmark the GAIA model, we have implemented a Google App Engine (GAE) based application that access geo-localized data in a testing datastore We compared the response time of each model to ours. It's important to mention that the size of the datastore will grow dynamically during the tests in a way that the results of the experiment will be based according to the growth of the datastore size, as well as compared with the increase of concurrent requests

In this benchmark, we have hosted the application in localhost, using the GAE local datastore that is implemented in "GAE for Eclipse" [8]. This is because we needed to test up to 10000 Queries Per Second (QPS), while the GAE trial version hosted online cannot accept more than 1 QPS. In addition, working with the actual GAE servers will only add network delay to our queries which does not affect our testing.

We populated 6 databases by generating random entries using the Poisson distribution [11], growing exponentially from 10 entries to 1M entries. Also, our tiny java web client, that generates requests to the server, can fork up to 10000 threads per second. Similarly, we have proceeded in an exponential scale, ranging from 1,10 to 10000 QPS.

In this testing we evaluated three criteria. These latter take into consideration the average time delay (ATD) as a basis for evaluation since the aim of this work is to improve response ATD and computation ATD for the cloud NRDBs. The criteria are:

- *Single queries evaluation (SQE):* This criterion evaluates the efficiency of one request relative to the growth of the data set size (DSS). For this, we will evaluate the data related to 1 QPS.
- *Concurrent queries evaluation (CQE):* This criterion evaluates the efficiency of the model under a stress of multiple parallel queries (which may retrieve the same data elements). This enables us to see if the model can be suited for efficient caching and indexing. In a real world scenario, the increase in QPS is directly related to the increase of the DSS. Hence, we are going to take the diagonal result that couples the increase of the number of requests with the increase of data-set.
- *Performance uniformity evaluation (PUE):* this criterion evaluates the uniformity of the performance of the model by analyzing its behavior under the strain on the variances on QPS and DSS. Unlike the previous evaluation that looks upon the performance variance according to the dataset/QPS relative change, this evaluation tests if the performance is quasi-constant, logarithmic, linear, linear, exponential, or quasi-random.

.

## V. Tests and results

By following the designed benchmark for the evaluation of GAIA, we have implemented modules in our back-end server that implements each of the models. The code was neither fine-tuned nor optimized to fit certain specifications of the platform. The platform is GAE local server with Java JPA enabled. The IDE for the testing used was Eclipse 3.7 Indigo [2]. The testing runs separately in a Windows desktop testing machine with all services and programs closed; only the application server and the requests client were running. The CPU of the machine has a dual core processor 1.86GHz and 2Go RAM. The testing run time took 16hours 27 minutes in total; the results were stored in log files, referring only to the tuples that holds the ATD, DSS, and QPS. The results of the tests are represented in the tables 1,2,3, and 4.

| ATD (sec) | Number of Parallel Requests (QPS) | | | | |
|---|---|---|---|---|---|
| Data Set Size (DSS) | 1 | 10 | 100 | 1000 | 10000 |
| 10 | 0.0011 | 0.0014 | 0.002 | 0.0201 | 0.05 |
| 100 | 0.0013 | 0.0016 | 0.0024 | 0.0199 | 0.0567 |
| 1000 | 0.0021 | 0.0033 | 0.0034 | 0.0343 | 0.0778 |
| 10000 | 0.0154 | 0.0233 | 0.0344 | 0.0731 | 0.1678 |
| 100000 | 0.0927 | 0.0999 | 0.1234 | 0.5678 | 1.237 |
| 1000000 | 0.3451 | 0.567 | 0.898 | 1.344 | 3.214 |

*Table 1 GRID method testing results representing the ATD with the variance of the QPS and DSS*

| ATD (sec) | Number of Parallel Requests (QPS) | | | | |
|---|---|---|---|---|---|
| Data Set Size (DSS) | 1 | 10 | 100 | 1000 | 10000 |
| 10 | 0.0011 | 0.0015 | 0.002 | 0.0191 | 0.0011 |
| 100 | 0.0012 | 0.0017 | 0.0023 | 0.0189 | 0.0012 |
| 1000 | 0.0018 | 0.0025 | 0.003 | 0.0243 | 0.0018 |
| 10000 | 0.0026 | 0.0031 | 0.0042 | 0.0354 | 0.0026 |
| 100000 | 0.003 | 0.0036 | 0.0045 | 0.039 | 0.003 |
| 1000000 | 0.0032 | 0.0039 | 0.0048 | 0.0471 | 0.0032 |

*Table 2 GAIA method testing results representing the ATD with the variance of the QPS and DSS*

| ATD (sec) | Number of Parallel Requests (QPS) | | | | |
|---|---|---|---|---|---|
| Data Set Size (DSS) | 1 | 10 | 100 | 1000 | 10000 |
| 10 | 0.0011 | 0.0015 | 0.002 | 0.0212 | 0.0675 |
| 100 | 0.0023 | 0.004 | 0.006 | 0.0195 | 0.0561 |
| 1000 | 0.0119 | 0.0137 | 0.0342 | 0.132 | 0.9873 |
| 10000 | 0.0454 | 0.0654 | 0.6456 | 0.9321 | 1.3454 |
| 100000 | 0.5173 | 0.677 | 0.991 | 1.5965 | 2.434 |
| 1000000 | 2.454 | 5.123 | 8.344 | 12.334 | 21.232 |

*Table 3 Projection method testing results representing the ATD with the variance of the QPS and DSS*

| ATD (sec) | Number of Parallel Requests (QPS) | | | | |
|---|---|---|---|---|---|
| Data Set Size (DSS) | 1 | 10 | 100 | 1000 | 10000 |
| 10 | 0.0014 | 0.0019 | 0.0021 | 0.021 | 0.0775 |
| 100 | 0.0023 | 0.0041 | 0.007 | 0.0191 | 0.0561 |
| 1000 | 0.0212 | 0.0237 | 0.0342 | 0.132 | 0.9873 |
| 10000 | 0.0521 | 0.0671 | 0.7329 | 1.124 | 1.3454 |
| 100000 | 0.6173 | 0.811 | 1.191 | 2.1865 | 3.176 |
| 1000000 | 3.154 | 7.213 | 13.349 | 24.274 | 51.213 |

*Table 4 Raw Coordinates method testing results representing the ATD with the variance of the QPS and DSS*

### V.1. Evaluation

We evaluated the tests results using the benchmark we have mentioned earlier, going through the SQE, CQE, and PUE, and we are providing a comparative table showcasing the bottom line of the benchmark.

<u>Single Query Evaluation:</u> Data resulted from the testing are represented in Tables 1 to 4, each cell represents the ATD related to a DSS and QPS. To have a meaningful insight, table 5 shows a comparison between the average time delays in each method, first the 1QPS ATD, and the ATD for large DSS. This shows the efficiency of the models in single queries, and under strain.

| Model | ATD for 1QPS (sec) | ATD for large DSS (sec) |
|---|---|---|
| RAW | 0.6414 | 48.4038 |
| Projection | 0.5053 | 20.2990 |
| GAIA | 0.0022 | 0.0898 |
| GRID | 0.0763 | 3.0222 |

*Table 5 : ATD for single query evaluation for the different Models*

From Table 5 we can see how GAIA outperforms the existing models in term of large dataset handling and simple single query handling. The Grid model remains the existing efficient model that is largely used by the industry. We can see also how raw coordinates handling and projection do not differ as much. Therefore, for the single query evaluation criterion, GAIA remains the most efficient model.

<u>Concurrent Queries Evaluation:</u> For this evaluation we plotted the ATDs that are related to diagonal results (DSS/QPS=10. This enabled us to have a representative sample from the results that can help us in this evaluation. The data in the Fig. 3 are plotted in a logarithmic scale. In the same figure we can see that the performance of GAIA does not significantly increase with the number of concurrent queries accessing the same data compared to the other models. This is mainly due to the thorough design requirements taken into consideration when developing the model.

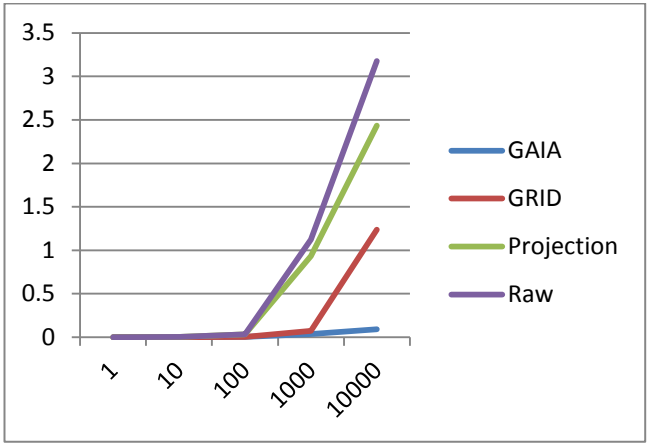

*Fig. 3 Plot of the ATD(y-axis) related to relative DSS (x-axis, or 10*QPS)*

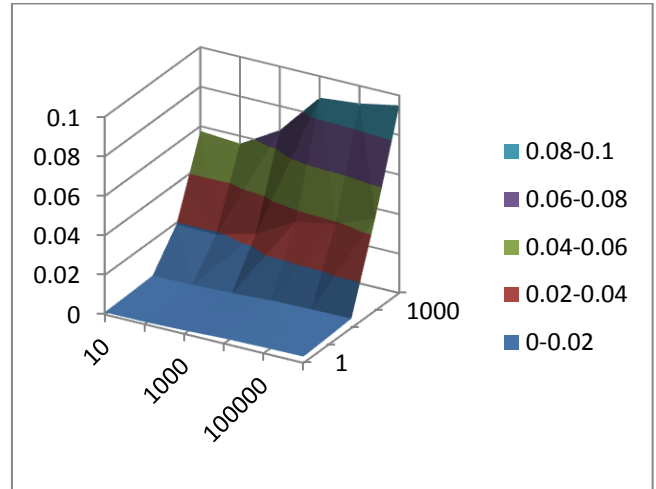

*Fig. 5 GAIA Method testing results plot of the ATD(z-axis) related to relative DSS (x-axis) and QPS (y-axis) in 0-0.1sec range*

Performance Uniformity Evaluation: As mentioned before, the PUE will enable us to evaluate and analyse the performance of the model based on the variance of the results ensued by the testing. For this, we depicted surface plots that represent a two variables function that relates ATD to both DSS and QPS. According to the previous results, the two most efficient models are the GRID model and the GAIA model. For the next comparisons, we are focusing mainly on the comparing GAIA to the GRID model, since the GRID model outperforms the existing models, which makes him a best fist for the study.

According to the regression results, in Fig. 6 and Fig. 7, the ATD can fit in the following approximated functions:

$$\mathbf{ADT_{GAIA}(DSS) = 0.000207211 * \log(10.1388 * DSS)} \quad (3)$$

$$\mathbf{ADT_{GRID}(DSS) = 0.138518 \cdot DSS + 0.1367} \quad (4)$$

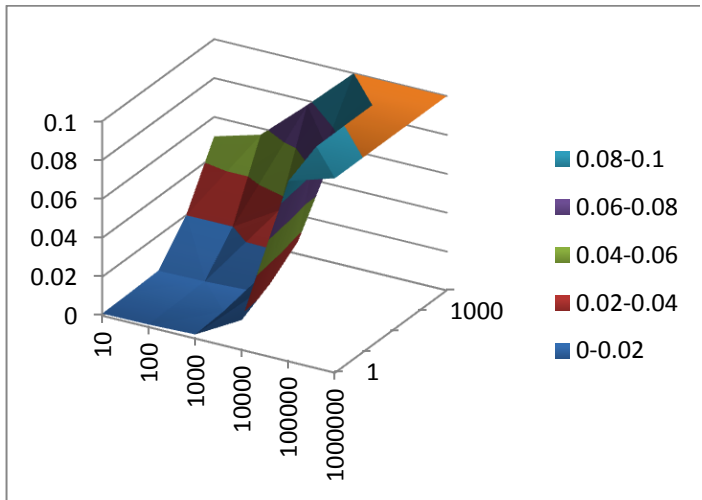

*Fig. 4 Grid Method testing results plot of the ATD(z-axis) related to relative DSS (x-axis) and QPS (y-axis) in 0-0.1sec range*

From the two plots in Fig. 4 and Fig. 5, we see that the change of the performance of GAIA is uniform while GRID has a linear growth. In other words, GAIA performance seems to be either logarithmic or bounded by a constant.

Using least square regression analysis, we compared the results of each model in order to approximate their efficiency, and make sure that our analysis is correct.

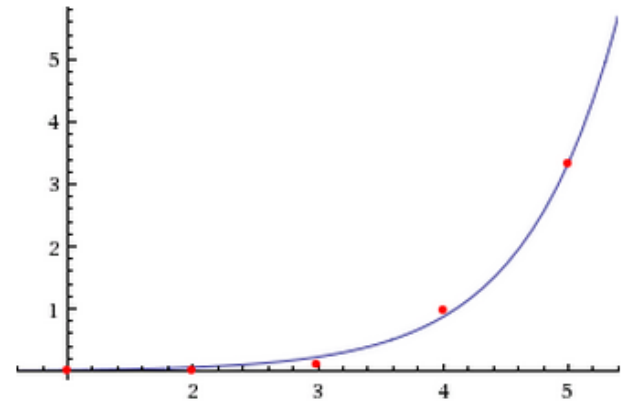

*Fig. 6 Regression using exponential functions for the GRID model (with a logarithmic scale for DSS at the x-axis, APT for y-axis)*

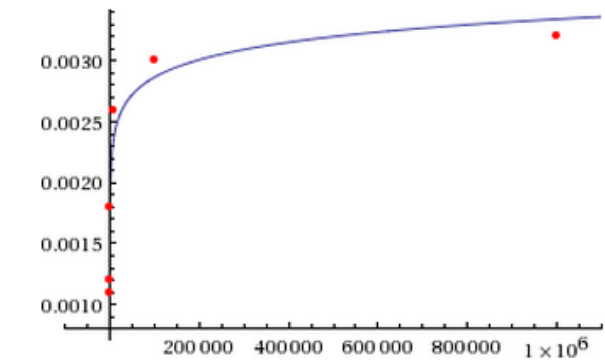

*Fig. 7 Regression using logarithmic functions for the GAIA model (with a linear scale for DSS at the x-axis, APT for y-axis)*

We may deduce that the performance of the GAIA algorithm (ADT) can be seen as logarithmically linear to DSS, while the existing models have linear efficiency.

Bottom Line: Assuming N is the number of rows in the table, S is the area of the surface to fetch, and c is the cell width. This table summarizes the benchmarking results:

|  | Projection | Coordinates | Grid | GAIA |
|---|---|---|---|---|
| Data labeling technique | One geometric coordinate | Geometric coordinates | Cell coordinates | Hash transformation function |
| Data Type | One Value Integer (1D) | Two Values Integers (2D) | Two Values Integers (2D) | One Value Integer (1D) |
| Method Used | Projection | None | Scaling | Scaling and Projection |
| Advantage | Hashing correlated data | precision | Precision, some correlation with cells | Precision and hashing correlated data |
| Disadvantage | Hashing non-correlated data | Data are not stored correlated | Based on the cells, not all data are precisely correlated | Requires some additional |
| SQE-ADT (sec) | 0.5053 | 0.6414 | 0.0763 | 0.0022 |
| CQE-ADT (sec) | 2.2748 | 2.599 | 1.1199 | 0.0852 |
| Data access complexity estimation from PUE | $O(\frac{N^2}{\sqrt{S}})$ | $O(N^2)$ | $O(\frac{N}{S*c})$ | $O(\log_{\frac{S}{c}}(N))$ |

*Table 6 Comparison of the existing models with GAIA using the designed benchmark*

## VI. Conclusion

We have studied alternative models for a spatially correlated data access, using non-relational databases (NRDB). Our goal is to minimize the average time delay (ADT) per query (QPS) for large data set sizes (DSS), and at the same time to engage as few database accesses as possible. To achieve these goals, we proposed the Geographically Asynchronous Information Access (GAIA) model, through:

- A simple mathematical transformation that combines the projection method and cell decomposition method (used for grids), which gave us a hash function G(p), that enables us to store the geographically correlated data (GCD) as neighbor NRDB segments.
- An algorithm for asynchronous and parallel access, designed to fit the nature of NRDBS in high replication databases (HRD), or any parallel architecture, to enable us retrieve correlated data simultaneously.
- A benchmark that allow us to test and compare the different existing models with GAIA, through following three evaluation access performance criteria.

The testing and comparison shows that GAIA outperforms the existing models, the Grid, the projection, and raw coordinate handling, since it provides a logarithmic efficiency at the order of O(log(n)), while the existing approaches have linear ADT. In future work, with an appropriate redundancy of cell size storage, data access efficiency can be reduced to a constant time.

The use of the GAIA method was basically designed to fetch non-relational databases that usually have row based access. Thus this approach was designed to answer this specific need. However, the method can be used with limited spatial data that can be better handled in one dimensional data structures or lists. Also, this method can improve Quad trees, R-trees, or R+trees for multidimensional data retrieval. Furthermore, the GAIA method was designed and implemented to be supported at the Application Layer (Software level), at the NDBMS level, and even to suit parallel computing architectures.

## VII. References

[1]. Khoabaltte, Ayoub. Jabeklah. www.jabeklah.com. [Online] 2012.
[2]. Retrieving and Indexing Spatial Data in the Cloud Computing Environment. Yonggang Wang, Sheng Wang and Daliang Zhou. 322-331, s.l. : CloudCom, Springer, 2009, Vol. 5931/2009.
[3]. Spherical Geodisc Grids. Heikes R., Randall D.A. I, Fort Collins : Monthly Weather Review, 1998, Vol. 123.
[4]. Interpolation on spherical geodesic grids: a. Carfora, Maria Francesca. 1-2, s.l. : Journal of Computational and Applied Mathematics, 2007, Vol. 210.
[5]. Pandey, Suraj. Cloud Computing Technology & GIS Applications. [PDF ] Melbourne : Department of Computer Science and Software Engineering, The University of Melbourne, Australia, 2010.
[6]. An Adaptive Grid Model for Urban to Regional Scale Air Quality Problems. Khan, M., T. Odman, H. Karimi and A. Hanna,. Istambul, Turkey : s.n., 2001. 2nd International Symposium on Air Quality Management at Urban.
[7]. OGC - About. Open Geospatial Consoritium. [Online] Open Geospatial Consoritium. http://www.opengeospatial.org/ogc.
[7]. Bigtable: A Distributed Storage System for Structured Data. Fay Chang, Jeffrey Dean, Sanjay Ghemawat, Wilson C. Hsieh. s.l. : OSDI2006: Seventh Symposium on Operating System Design and Implementation, 2006.
[9]. Intercomparison of Spatial Interpolation Schemes for Use in Nested Grid Models. Kiran Alapaty, Rohit Mathur, and Talat Odman. I, s.l. : Monthly Weather Review, 1998, Vol. 126.
[10]. Conos, Michal. Recognition of vehicle make from a frontal view. [PDF] s.l. : VELLUM Ltd, 3 01, 2007.

[11]. chang, K.-t. Introduction to Geographic Information Systems. New York : McGraw-Hill Companies, Inc, 2002.

[12]. Johnson, Don H. Image Projections and the Radon Transform. ELEC 431: Digital Signal Processing. [Online] Curricular Linux Environment at Rice University (CLEAR), Houston, Texas, 2 2006, 5. [Cited: 6 2012, 16.] http://www.clear.rice.edu/elec431/projects96/DSP/bpanalysis.html.